\documentclass[11pt,a4paper]{article}
\usepackage[truedimen,margin=34mm]{geometry} 

\usepackage{natbib}
\usepackage{mathrsfs}
\usepackage{amssymb}
\usepackage{amsmath}
\usepackage{ascmac}
\usepackage{amsthm}
\usepackage[pdftex]{graphicx}
\usepackage[pdftex]{color}
\usepackage{booktabs}
\usepackage{setspace}
\usepackage{color}

\newtheorem{algo}{Algorithm}

\def\ep{{\varepsilon}}
\def\om{{\omega}}
\def\si{{\sigma}}
\def\th{{\theta}}

\def\rhoh{{\widehat \rho}}

\def\psih{{\widehat \psi}}
\def\tah{{\widehat \tau}}
\def\muh{{\widehat \mu}}

\def\sh{\widehat{s}}
\def\omh{\widehat{\omega}}

\def\Si{{\Sigma}}

\def\diag{{\rm diag}}
\def\tr{{\rm tr}}

\def\E{{\rm E}}

\title{{\bf A unified method for improved inference in random-effects meta-analysis}}
\date{}

\begin{document}

\doublespacing
\maketitle

\vspace{-1.5cm}

\begin{center}
SHONOSUKE SUGASAWA\\[4pt]
\textit{Center for Spatial Information Science, The University of Tokyo, 5-1-5, Kashiwanoha, Kashiwa, Chiba, Japan}
\\[2pt]
{sugasawa@csis.u-tokyo.ac.jp}\\[6pt]
HISASHI NOMA\\[4pt]
\textit{Research Center for Medical and Health Data Science, The Institute of
Statistical Mathematics, 10-3, Midori-cho, Tachikawa, Tokyo, Japan
}
\end{center}

\vspace{0.3cm}
\begin{abstract}
{
Random-effects meta-analyses have been widely applied in evidence synthesis for various types of medical studies. However, standard inference methods (e.g. restricted maximum likelihood estimation) usually underestimate statistical errors and possibly provide highly overconfident results under realistic situations; for instance, coverage probabilities of confidence intervals can be substantially below the nominal level. The main reason is that these inference methods rely on large sample approximations even though the number of synthesized studies is usually small or moderate in practice. In this article we solve this problem using a unified inference method based on Monte Carlo conditioning for broad application to random-effects meta-analysis. The developed method provides improved confidence intervals with coverage probabilities that are closer to the nominal level than standard methods. As specific applications, we provide new inference procedures for three types of meta-analysis: conventional univariate meta-analysis for pairwise treatment comparisons, meta-analysis of diagnostic test accuracy, and multiple treatment comparisons via network meta-analysis. We also illustrate the practical effectiveness of these methods via real data applications and simulation studies.
}
{
Confidence interval; Meta-analysis; Likelihood ratio test; Random-effects model
}
\end{abstract}

\section{Introduction}
\label{sec1}

In evidence-based medicine, meta-analysis has been an essential tool for quantitatively summarizing multiple studies and producing integrated evidence. 
In general, the treatment effects from different sources of evidence are heterogeneous due to various factors, which should be adequately addressed, otherwise statistical errors may be seriously underestimated and possibly result in misleading conclusions \citep{Higgins2011}. 
Such heterogeneity can be generally divided into two types, fixed-effects and random-effects, and the comparison between two methods have been discussed \citep[e.g.][]{Hedges1998,Overton1998}.
As noted in \cite{Rice2018}, fixed-effects and random-effects methods are respectively validated under different settings and assumptions.
On the other hand, random-effects models are widely used in most medical meta-analyses.
The applications cover various types of systematic reviews, for example, conventional univariate meta-analysis \citep{DL1986,WW1991}, meta-analysis of diagnostic test accuracy \citep{Reitsma2005}, network meta-analysis for comparing the effectiveness of multiple treatments \citep{Salanti2012}, and individual participant meta-analysis \citep{Riley2010}. 
In view of this background, we focused on random-effects meta-analysis in this paper.

However, in random-effects meta-analysis, most existing standard inference methods (e.g. restricted maximum likelihood method) for average treatment effect parameters underestimate statistical errors under realistic situations of medical meta-analysis.
For example, the coverage probabilities of standard inference methods are usually smaller than the nominal confidence levels, even when the model is completely specified \citep{BG2001}, which may lead to highly overconfident conclusions. 
This notable problem is related to heterogeneity variance-covariance parameters, typically treated as nuisance parameters, included in random-effects models. 
When the number of studies is small, estimators (e.g. restricted maximum likelihood estimator) of such variance parameters have high variability, so that they tend to underestimate the true value and sometimes produce an exact estimate of 0.
In the standard inference methods, such variability is often ignored using large sample approximations for the number of studies whereas the number is small or moderate in medical meta-analysis, which results that the total statistical error can be underestimated when constructing confidence intervals.
So far, several confidence intervals that aim to improve the undercoverage property have been developed, for example, by \cite{HK2001a}, \cite{HK2001b}, \cite{Henmi2010}, \cite{Jackson2009}, \cite{Jackson2014}, \cite{KH2003}, \cite{Noma2011}, \cite{Noma2018}, \cite{Guolo2012}, \cite{SM2008} and \cite{Sidik2002}. 
Although these methods improve coverage properties, they are substantially valid under the large number of samples.
Moreover, most of these methods were developed in the context of traditional direct pairwise comparisons; therefore, the methods have limited applicability in recent, more advanced types of meta-analysis that use the complicated multivariate models.

In this paper, we develop a unified method for constructing confidence intervals (regions) for parameters in random-effects models so that their coverage probabilities are almost equal to the nominal level regardless of the number of studies. 
To effectively circumvent the effects of nuisance parameters, we consider the likelihood ratio test (LRT) for the average treatment effect, and we define its $p$-value based on the conditional distribution given the maximum likelihood estimator of the nuisance parameters rather than the unconditional distribution of the test statistic. 
For computing the $p$-value, we adopt the Monte Carlo conditioning technique proposed by \cite{LT2005}, and the confidence interval can be derived by inverting the LRT. 
As a result, the derived confidence intervals are shown to have reasonable coverage probabilities and the proposed method can be generally applied to various types of meta-analysis involving complicated multivariate random-effects models.

This paper is organized as follows.
In Section \ref{sec:general}, we first provide an algorithm for computing the $p$-value of the LRT and derive confidence intervals under general statistical models. 
In Sections \ref{sec:uni}, \ref{sec:DMA} and \ref{sec:NMA}, we demonstrate the proposed method in univariate meta-analysis for direct pairwise comparisons, bivariate meta-analysis for diagnostic test accuracy and network meta-analysis, respectively, using real datasets and simulations. 
Discussion is provided in Section \ref{sec:Disc}.

\section{Algorithm for confidence interval}\label{sec:general}

We suppose $y_1,\ldots,y_k$ are independent and each has the density or probability mass function $f_i(y_i; \phi, \psi)$ with parameter of interest $\phi$ and nuisance parameter $\psi$.
For example, in the univariate meta-analysis described in Section \ref{sec:uni}, $k$ and $y_i$ correspond to the number of studies and estimated treatment effect in the $i$th study, respectively, and we use the model $y_i\sim N(\mu,\tau^2+\si_i^2)$ with known $\si_i^2$ to estimate the average treatment effect $\mu$, so that $\phi=\mu$ and $\psi=\tau^2$ in this case. 
In general, $y_i$, $\phi$ and $\psi$ could be multivariate, but we assume in this section that all of them are one-dimensional in order to make our presentation simpler.
Multivariate cases are considered in Sections \ref{sec:DMA} and \ref{sec:NMA}.
The likelihood ratio test (LRT) statistic for testing null hypothesis $H_0: \phi=\phi_0$ is 
$T_{\phi_0}(Y)=-2\left\{\max_{\psi}L(Y,\phi_0,\psi)-\max_{\phi,\psi}L(Y,\phi,\psi)\right\}$,
where $Y=(y_1,\ldots,y_k)^t$, and $L(Y,\phi,\psi)=\sum_{i=1}^k\log f_i(y_i; \phi, \psi)$ is the log-likelihood function.
Under some regularity conditions, the asymptotic distribution of $T_{\phi_0}(Y)$ under $H_0$ is $\chi^2(1)$ as $k\to\infty$.
However, when the sample size $k$ is not large as is often the case in meta-analysis, the approximation is not accurate enough.
The main reason is that there is an unknown nuisance parameter $\psi$ and its estimation error is not ignorable when $k$ is not large.
To overcome this problem, we calculate the $p$-value of the statistic $T_{\phi_0}(Y)$ based on the conditional distribution $Y|\psih(\phi_0)$, where $\psih(\phi_0)$ is the maximum likelihood estimator of $\psi$ under $H_0$.
For computing the $p$-value, we adopt the Monte Carlo conditioning developed in \cite{LT2005}.

To describe the general methodology, we further assume that $Y$ can be expressed as $Y=H(U,\phi,\psi)$ for some function $H$ and random variable $U$ whose distribution is completely known.
For example, when $Y\sim N(\phi,\psi)$, it holds that $Y=\phi+\sqrt{\psi} U$ with $U\sim N(0,1)$.
Now, the maximum likelihood estimator $\psih$ under $H_0$ satisfies the equation, $L_{\psi}(Y,\phi_0,\psih)=0$, where $L_{\psi}=\partial L/\partial\psi$ is the partial derivative of the likelihood function $L$ with respect to the nuisance parameter $\psi$.
Under $H_0$, $Y$ can be expressed as $Y=H(U,\phi_0,\psi)$; thereby, the above equation can be rewritten as 
$$
\delta(U,\psih,\psi)\equiv L_{\psi}(H(U,\phi_0,\psi),\phi_0,\psih)=0.
$$
We define $\psi_{\ast}(U)$ as the solution of the above equation with respect to $\psi$.
Using the result in \cite{LT2005}, the $p$-value $\E\big[I\{T(Y)\geq t\}|\psih\big]$ can be expressed as 
\begin{equation}\label{pval}
\E\big[I\{T(Y)\geq t\}|\psih\big]=\frac{\E\big[I\{T(H(U,\phi_0,\psi_{\ast}(U)))\geq t\}w(U)\big]}{\E\big[w(U)\big]},
\end{equation}
where the expectation is taken with respect to the distribution of $U$, and 
\begin{equation}\label{weight}
w(U)=\Bigg|\frac{\pi(\psi)}{\partial\psih/\partial\psi}\Bigg|_{\psi=\psi_{\ast}(U)}
\end{equation}
with some function $\pi(\psi)$ of $\psi$.
As noted in \cite{LT2005}, the choice of $\pi(\psi)$ controls the efficiency of the Monte Carlo approximation in (\ref{pval}).
However, the detailed discussion of this issue would extend of the scope of this paper; thus, we consider in this paper only $\pi(\psi)=1$.
The algorithm for computing the $p$-value for testing H$_0: \phi=\phi_0$ is given as follows.

\begin{algo}\label{algo:p}
(Monte Carlo method for $p$-value of LRT)
\begin{itemize}
\item[1.]
For $b=1,\ldots,B$ with large $B$, generate a random sample, $U^{(b)}=(u_1^{(b)},\ldots,u_k^{(b)})$, and compute $\psi_{\ast}(U^{(b)})$, $Y_{\ast}^{(b)}=H(U^{(b)},\phi_0,\psi_{\ast}(U^{(b)}))$ and $w(U^{(b)})$ from (\ref{weight}).

\item[2.]
The Monte Carlo approximation of the $p$-value is given by 
\begin{equation*}
\frac{\sum_{b=1}^BI\left\{T_{\phi_0}(Y_{\ast}^{(b)})\geq T_{\phi_0}(Y)\right\}w(U^{(b)})}{\sum_{b=1}^Bw(U^{(b)})}.
\end{equation*}
\end{itemize}
\end{algo}

Using the $p$-value of the LRT of $H_0: \phi=\phi_0$, the confidence interval of $\phi$ with nominal level $1-\alpha$ can be constructed as the set of $\phi^{\dagger}$ such that the $p$-value of the LRT of $H_0: \phi=\phi^{\dagger}$ is larger than $\alpha$. 
Although the confidence limits cannot be expressed in closed form, they can be computed by simple numerical methods, for example the bisectional method that repeatedly bisects an interval and selects a subinterval in which a root exists until the process converges numerically, see Section 2 in \cite{Burden2010}.
When $\psi$ is multivariate (vector-valued) parameters, which is typical in many applications, the absolute value symbol in the weight (\ref{weight}) should be recognized as the absolute value of determinant since $\partial\psih/\partial\psi^t$ is a matrix.
On the other hand, when $\phi$ is multivariate, we need to construct a confidence region (CR) rather than a confidence interval. 
In this case, the bisectional method cannot be directly applied, and methods for CRs would depend on each setting.  
In Section \ref{sec:DMA}, we present a diagnostic meta-analysis in which a CR is traditionally used, and provide a feasible algorithm to compute a CR.

\section{Univariate random-effects meta-analysis}\label{sec:uni}

\subsection{The random-effects model}
The univariate random-effects model has been widely used in meta-analysis due to its parametric simplicity.
However, the accuracy of the inference is poor when the number of studies is small.
We consider solving this problem using the LRT and confidence interval introduced in Section \ref{sec:general}. 
We assume that there are $k$ clinical trials and that $y_1,\ldots,y_k$ are the estimated treatment effects.
We consider the random-effect model:
\begin{equation}\label{uni-model}
y_i=\th_i+e_i, \ \ \ \ \th_i=\mu+\ep_i, \ \ \ i=1,\ldots,k,
\end{equation}
where $\th_i$ is the true effect size of the $i$th study, and $\mu$ is the average treatment effect. 
Here $e_i$ and $\ep_i$ are independent error terms within and across studies, respectively, assumed to be distributed as $e_i\sim N(0,\si_i^2)$ and $\ep_i\sim N(0,\tau^2)$. 
The within-studies variances $\si_i^2$s are usually assumed to be known and fixed to their valid estimates calculated from each study. 
On the other hand, the across variance $\tau^2$ is an unknown parameter representing the heterogeneity between studies. 
Under these settings, \cite{HT1996} considered the likelihood-based approach for estimating the average treatment effect $\mu$.

\subsection{Confidence intervals of model parameters}
We first consider a confidence interval of $\mu$ by using Algorithm 1 in the previous section.
To begin with, we consider the null hypothesis $H_0: \mu=\mu_0$ with nuisance parameter $\tau^2$.
Since $y_i\sim N(\mu,\tau^2+\si_i^2)$ under the model (\ref{uni-model}), the LRT statistic can be defined as 
$T_{\mu_0}(Y)=\min_{\mu,\tau^2}L(\mu,\tau^2)-\min_{\tau^2}L(\mu_0,\tau^2)$,
where
\begin{equation}\label{uni-like}
L(\mu,\tau^2)=\sum_{i=1}^k\log(\tau^2+\si_i^2)+\sum_{i=1}^k\frac{(y_i-\mu)^2}{\tau^2+\si_i^2}.
\end{equation}
The minimization can be achieved in standard ways such as the iterative method in \cite{HT1996}.

From (\ref{uni-like}), the constrained maximum likelihood estimator $\tah_c^2$ of $\tau^2$ under H$_0$ satisfies the following equation: 
$$
\sum_{i=1}^k\frac{1}{\tah_c^2+\si_i^2}-\sum_{i=1}^k\frac{(y_i-\mu_0)^2}{(\tah_c^2+\si_i^2)^2}=0.
$$
Let $u_1,\ldots,u_m$ be random variables that which independently follow $N(0,1)$, then $y_i$ can be expressed as $y_i=\mu_0+u_i\sqrt{\tau^2+\si_i^2}$ under H$_0$.
Substituting the expression for $y_i$ in the above equation, we have
\begin{equation*}
G(U,\tau^2,\tah_c^2)\equiv \sum_{i=1}^k\frac{1}{\tah_c^2+\si_i^2}-\sum_{i=1}^k\frac{(\tau^2+\si_i^2)u_i^2}{(\tah_c^2+\si_i^2)^2}=0,
\end{equation*}
where $U=(u,\ldots,u_k)^t$.
The above equation can be easily solved with respect to $\tau^2$ and the solution is given by
$$
\tau_{\ast}^2(U)=\left\{\sum_{i=1}^k\frac{u_i^2}{(\tah_c^2+\si_i^2)^2}\right\}^{-1}\left\{\sum_{i=1}^k\frac{\tah_c^2+\si_i^2(1-u_i^2)}{(\tah_c^2+\si_i^2)^2}\right\}.
$$
Regarding the weight (\ref{weight}), using the implicit function theorem for the equation $G(U,\tau^2,\tah_c^2)=0$, it holds that 
\begin{align*}
w(U)
=\left\{\sum_{i=1}^k\frac{u_i^2}{(\tah_c^2+\si_i^2)^2}\right\}^{-1}\left|\sum_{i=1}^k\frac{2\{\tau_{\ast}^2(U)+\si_i^2\}u_i^2-(\tah_c^2+\si_i^2)}{(\tah_c^2+\si_i^2)^3}\right|,
\end{align*}
and thereby we can compute the $p$-value of the LRT for H$_0: \mu=\mu_0$ from Algorithm 1, and the confidence interval of $\mu$ by inverting the LRT.

A confidence interval of $\tau^2$ can be derived as well.
By a similar derivation to that for $\mu$, the $p$-values of the LRT of H$_0: \tau^2=\tau_0^2$ can be computed from Algorithm 1 with $w(U)=1$ and  
$$
\mu_{\ast}(U)=\muh-\left(\sum_{i=1}^k\frac{1}{\tau_0^2+\si_i^2}\right)^{-1}\sum_{i=1}^k\frac{u_i}{\sqrt{\tau_0^2+\si_i^2}},
$$
so that the confidence interval of $\tau^2$ can be similarly constructed.

\subsection{Simulation study}\label{sec:sim}
We evaluated the finite sample performance of the proposed confidence interval of $\mu$ via Monte Carlo (MC) algorithm together with existing methods widely used in practice.
We considered the restricted maximum likelihood (REML) method, the DerSimonian and Laird (DL) method \citep{DL1986}, the Knapp and Hartung (KNHA) method \citep{KH2003} with random-effects variance estimated by REML, and the likelihood ratio (LR) method \citep{HT1996}.
When implementing the proposed MC method, we used $1000$ Monte Carlo samples to compute the $p$-value.
We fixed the true average treatment effect $\mu$ at $-0.80$, and the heterogeneity variance $\tau^2$ at $0.10$ and $0.20$.
We changed the number of studies $k$ over $3,5,7$ and $9$, and set the nominal level $\alpha$ to $0.05$.
To approximate practical situations of medical meta-analyses, we followed the simulation settings considered by \cite{Sidik2007}.
We generated $\theta_i\sim N(\mu,\tau^2)$ and binomial data $X_{ir}\sim {\rm Binomial}(n_{ir},p_{ir})$ for $i=1,\ldots,k$ and $r=0,1$ corresponding to control and treatment. 
The response rate of control $p_{i0}$ was generated from a continuous uniform distribution on $[0.095,0.65]$ and we set $p_{i1}=p_{i0}\exp(\th_i)/\{1-p_{i0}+p_{i0}\exp(\theta_i)\}$, which means that $\theta_i$ is odds ratio, i.e. $\theta_i={\rm legit}(p_{i1})-{\rm legit}(p_{i0})$.
The sample sizes were set to equal $n_{i1}=n_{i0}$ and were randomly sampled with replacement from the integers between 20 and 200.
For the simulated binomial data, the summary statistics for $\theta_1,\ldots,\theta_k$, i.e. sample log odds ratios $y_i$ and their estimated asymptotic variances $s_i^2$, are calculated. 
Based on $2000$ simulation runs, we calculated the coverage probabilities (CP) and average lengths (AL) of the four confidence intervals.

The results, shown in Table \ref{tab:CP}, indicate that the confidence intervals from the three methods, LR, REML and DL tend to be liberal to  achieve the appropriate nominal level $0.95$ even when $k=9$.
On the other hand, the proposed MC method produces reasonable confidence intervals with appropriate coverage probabilities even when $k$ is $3$.
KNHA also provides reasonable confidence intervals even when $k=3$, but overall it tends to be slightly liberal compared with MC.
The average lengths of MC and KNHA are similar and longer than those of the other methods.

\begin{table}[!htb]
\caption{Simulated coverage probabilities (\%) and average lengths for $95\%$ confidence intervals from the proposed Monte Carlo (MC) method, the method by \cite{KH2003}, the likelihood ratio (LR) method, the restricted maximum likelihood (REML) method, and the DerSimonian and Laird (DL) method.
\label{tab:CP}
}
\centering
\vspace{0.5cm}
\begin{tabular}{ccccccccccccccccccc}
\hline
&& \multicolumn{4}{c}{Coverage Probability (\%)} && \multicolumn{4}{c}{Average Length}\\
  & $k$ & 3 & 5 & 7 & 9 &  & 3 & 5 & 7 & 9\\
\hline
& MC & 96.6 & 96.1 & 96.4 & 95.3 &  & 2.032 & 1.146 & 0.861 & 0.710 \\
 & KNHA & 93.6 & 94.7 & 94.6 & 93.8 &  & 2.097 & 1.090 & 0.823 & 0.686 \\
$\tau^2=0.10$ & LR & 92.8 & 93.7 & 93.5 & 92.6 &  & 1.233 & 0.884 & 0.725 & 0.626 \\
 & REML & 88.9 & 91.5 & 91.6 & 91.1 &  & 1.064 & 0.801 & 0.673 & 0.589 \\
 & DL & 89.2 & 91.8 & 92.0 & 90.8 &  & 1.068 & 0.801 & 0.673 & 0.590 \\
 \hline
 & MC & 94.7 & 95.7 & 95.4 & 95.2 &  & 2.360 & 1.356 & 1.033 & 0.862 \\
 & KNHA & 93.5 & 94.4 & 94.9 & 94.5 &  & 2.482 & 1.310 & 1.005 & 0.843 \\
$\tau^2=0.20$ & LR & 88.8 & 91.4 & 92.2 & 93.7 &  & 1.396 & 1.039 & 0.869 & 0.759 \\
 & REML & 83.9 & 89.2 & 90.3 & 92.1 &  & 1.207 & 0.941 & 0.805 & 0.715 \\
 & DL & 84.6 & 89.0 & 90.9 & 92.1 &  & 1.205 & 0.939 & 0.804 & 0.713 \\
\hline
\end{tabular}
\end{table}

\subsection{Example: treatment of suspected acute myocardial infarction}\label{sec:mag}
Here we applied the proposed method to a meta-analysis of the treatment of suspected acute myocardial infarction with intravenous magnesium \citep{Teo1991}, which is well-known as it yielded conflicting results between meta-analyses and large clinical trials \citep{LeL1997}. 
For the dataset, we constructed a $95\%$ confidence interval of the average treatment effect of intravenous magnesium using the proposed MC method (with $10000$ Monte Carlo samples) as well as the KNHA, LR, REML and DL methods considered in the previous section.
Moreover, we also applied Peto's fixed effect (PFE) method \citep{Yu1985}.
The detailed results are given in Supplementary Material.
We found that the confidence intervals from the KNHA, DL, LR, REML and PFE methods were narrower than that of the MC method, and the confidence intervals from the five methods did not cover $\mu=1$, which does not change the interpretation of the results. 
On the other hand, the proposed MC method produced a longer confidence interval while also covering $\mu=1$, that is, the corresponding test for $\mu=1$ was not significant with a $5\%$ significant level.


\section{Bivariate Meta-analysis of Diagnostic Test Accuracy}\label{sec:DMA}

\subsection{Bivariate random-effects model}

There has been increasing interest in systematic reviews and meta-analyses of data from diagnostic accuracy studies.
For this purpose, a bivariate random-effect model \citep{Reitsma2005, Harbord2007} is widely used.
Following \cite{Reitsma2005}, we define $\mu_{Ai}$ and $\mu_{Bi}$ as the logit-transformed true sensitivity and specificity, respectively, in the $i$th study. 
The bivariate model assumes that $(\mu_{Ai},\mu_{Bi})^t$ follows a bivariate normal distribution:
\begin{equation}\label{BMA1}
\left(\begin{array}{c} \mu_{Ai} \\ \mu_{Bi} \end{array}\right)
\sim
N\left(\left(\begin{array}{c} \mu_{A} \\ \mu_{B} \end{array}\right), \Si\right) \ \ \ \ 
\text{with}\ \ \ 
\Si=\left(\begin{array}{cc} \si_A^2  & \rho\si_A\si_B\\ \rho\si_A\si_B & \si_B^2 \end{array}\right),
\end{equation}
where $\mu_A$ and $\mu_B$ are the average logit-transformed sensitivity and specificity, and $\si_A(>0)$ and $\si_B(>0)$ are standard deviations of $\mu_{Ai}$ and $\mu_{Bi}$, respectively.
Here the parameter $\rho\in (-1,1)$ allows correlation between $\mu_{Ai}$ and $\mu_{Bi}$.
The unknown parameters are $\mu_A,\mu_B,\si_A^2,\si_B^2$ and $\rho$.
Let $y_{Ai}$ and $y_{Bi}$ be the observed logit-transformed sensitivity and specificity, and we assume that  
\begin{equation}\label{BMA2}
\left(\begin{array}{c} y_{Ai} \\ y_{Bi} \end{array}\right)
\sim
N\left(\left(\begin{array}{c} \mu_{Ai} \\ \mu_{Bi} \end{array}\right), C_i\right) \ \ \ \ 
\text{with}\ \ \ 
C_i=\left(\begin{array}{cc} s_{Ai}^2  & 0\\ 0 & s_{Bi}^2 \end{array}\right).
\end{equation}
For summarizing the results of the meta-analysis, the CR of $\mu=(\mu_A,\mu_B)^t$ would be more valuable than separate confidence intervals since sensitivity and specificity might be highly correlated.
\cite{Reitsma2005} suggested the $100(1-\alpha)\%$ joint CR for $\mu$ as the interior points of the ellipse defined as
\begin{equation}\label{NCR}
\mu_A=\muh_A+c_{\alpha}\sh_A\cos t, \ \ \ \ 
\mu_B=\muh_B+c_{\alpha}\sh_B\cos(t+\arccos \rhoh), \ \ \ \ t\in [0,2\pi),
\end{equation}
where $\muh_A$ and $\muh_B$ are estimates of $\mu_A$ and $\mu_B$, $\sh_A$ and $\sh_B$ are estimated standard errors of $\muh_A$ and $\muh_B$, respectively, which are obtained via the restricted maximum likelihood method.
Here $c_{\alpha}$ is the square root of the upper $100\alpha\%$ point of the $\chi^2$ distribution with $2$ degrees of freedom.
The joint CR (\ref{NCR}) is approximately valid; specifically, the coverage error converges to $1-\alpha$ as the number of studies $k$ goes to infinity.
However, when $k$ is not sufficiently large, the coverage error is not negligible, and the region (\ref{NCR}) would under-cover the true $\mu$.

\subsection{Confidence region of sensitivity and specificity}

We consider a CR of $\mu$ under the models (\ref{BMA1}) and (\ref{BMA2}) based on the Monte Carlo method given in Section \ref{sec:general}.
Let $\psi=(\si_A^2,\si_B^2,\rho)^t$ be a vector of nuisance parameters, and write $\Si(\psi)$ rather than $\Si$ to clarify the dependence of $\psi$.  
From (\ref{BMA1}) and (\ref{BMA2}), the LRT statistic of the null hypothesis H$_0: \mu=\mu_0$ is given by  
$$
T_{\mu_0}(Y)=\min_{\mu,\psi}L(\mu,\psi)-\min_{\psi}L(\mu_0,\psi),
$$
where
$$
L(\mu,\psi)=\sum_{i=1}^k\log|V_i(\psi)|+\sum_{i=1}^k(y_i-\mu)^tV_i(\psi)^{-1}(y_i-\mu),
$$
with $V_i(\psi)=\Si(\psi)+C_i$.

Under H$_0$, the constrained maximum likelihood estimator $\psih_c$ satisfies the following equations:
$$
\sum_{i=1}^k\tr\{V_i(\psih_c)^{-1}J_k\}-\sum_{i=1}^k(y_i-\mu_0)^tV_i(\psih_c)^{-1}J_kV_i(\psih_c)^{-1}(y_i-\mu)=0,  \ \ \ \ k=1,2,3,
$$
where
$$
J_1=\left(
\begin{array}{cc}
1 & 0 \\
0 & 0
\end{array}
\right), \ \ \ \ 
J_2=\left(
\begin{array}{cc}
0 & 0 \\
0 & 1
\end{array}
\right), \ \ \ \ 
J_3=\left(
\begin{array}{cc}
0 & 1 \\
1 & 0
\end{array}
\right). 
$$
Under H$_0$, the observed data $y_i$ can be expressed as $y_i=\mu_0+T_i(\psi)u_i$ with $u_i\sim N(0,I_2)$ and $T_i(\psi)$ being the Cholesky decomposition of $V_i(\psi)$, that is, $T_i(\psi)T_i(\psi)^t=V_i(\psi)$.
Then the above equation can be rewritten as 
\begin{equation*}\label{BMA-eq}
\sum_{i=1}^k\tr\{V_i(\psih_c)^{-1}J_k\}-\sum_{i=1}^ku_i^tT_i(\psi)^tV_i(\psih_c)^{-1}J_kV_i(\psih_c)^{-1}T_i(\psi)u_i=0,  \ \ \ \ k=1,2,3,
\end{equation*}
and we define $\psi_{\ast}(U)$ as the solution of the above equation with respect to $\psi$.
The solution can be numerically obtained by minimizing the sum of squared values of three equations with respect to $\psi$.
Moreover, concerning the weight (\ref{weight}), we can use the numerical derivative given $U$ to compute the partial derivative $\partial\psih/\partial \psi^t$ evaluated at $\psi=\psi_{\ast}(U)$.
Hence, we can compute the $p$-value of the LRT of H$_0:\mu=\mu_0$ from Algorithm \ref{algo:p}.

Using the LRT of H$_0$, the $(1-\alpha)\%$ CR of $\mu$ can be defined as $\text{ECR}_{\alpha}=\{\mu ; p(\mu)\geq \alpha\}$, where $p(\mu)$ denotes the $p$-value of the test statistic $T_{\mu}(Y)$.
Since $\mu$ is two-dimensional in this case, the computing boundary $\{\mu ; p(\mu)=\alpha\}$ is not straightforward.
The most feasible procedure is to approximate the boundary with a sufficiently large numbers of points.
To this end, we first divide the interval $[0,2\pi)$ by $M$ points $0=t_1<\cdots <t_M<2\pi$.
For each $m=1,\ldots,M$, we compute $r_m$ satisfying 
$$
p(\muh+(r_m\cos t_m,r_m\sin t_m))=\alpha,
$$
which can be carried out via numerical methods (e.g. the bisectional method).

\subsection{Simulation study}\label{sec:BMA-sim}
We assessed the finite sample performance of the proposed confidence region via Monte Carlo (MC) algorithm together with the approximate confidence region (\ref{NCR}) by \cite{Reitsma2005}.
When implementing the MC method, we used 500 Monte Carlo samples to compute the $p$-value.
We set $\mu_A=1, \mu_B=-1$ and $\tau_A=\tau_B(=\tau)$.
We used the between study variances $\tau^2$ of $0.5$, $0.75$ and $1$, and the between study correlations $\rho$ of $0, 0.4$ and $0.8$.
Following, \cite{Jackson2014}, for each simulation, two sets of $k$ within-study variances were simulated from a scaled chi-squared distribution with 1 degree of freedom, multiplied by 0.25, and truncated to lie within the interval $[0.009, 0.6]$.
We changed the number of studies $k$ over 8,12 and 16, and set the nominal level $\alpha$ to $0.05$.
In the 1000 simulations, we evaluated empirical coverage probabilities for 95\% confidence regions of the true parameters.
Since the MC method requires unrealistic computational times to calculate boundaries of a confidence region in large simulations (one such calculation can be implementable within a reasonable time), we only evaluated coverage rates assessing rejection rates of the test of null hypothesis for the true parameters. 
The concrete example of the confidence region based on the MC method is illustrated in Section \ref{sec:BMA-app}.

The results of the simulations are shown in Table \ref{tab:BMA-sim}.
The simulated coverage probabilities of the standard method, ACR, are seriously smaller than the nominal level ($95\%$), especially in the case with the small number of studies ($k=8$).
Such undesirable results would come from the crude approximation in (\ref{NCR}).
On the other hand, the simulated coverage probabilities of the proposed MC method are around the nominal level in all the scenarios, as expected, which indicates that the proposed method can produce reasonable confidence region with adequate assessment of the statistical error of the estimation of $(\mu_A,\mu_B)$.

\begin{table}[!htb]
\caption{Simulated coverage probabilities (\%) for $95\%$ confidence regions from the proposed Monte Carlo (MC) method, and the approximated (ACR) method by \cite{Reitsma2005}. 
\label{tab:BMA-sim}
}
\centering
\vspace{0.5cm}
\begin{tabular}{ccccccccccccccccccc}
\hline
&&& \multicolumn{2}{c}{$\rho=0$} & \multicolumn{2}{c}{$\rho=0.4$} &\multicolumn{2}{c}{$\rho=0.8$}\\
 & $k$ && MC & ACR & MC & ACR & MC & ACR \\
\hline
& 8 &  & 94.2 & 77.4 & 94.4 & 75.9 & 94.7 & 75.5\\
$\tau^2=0.5$ & 12 &  & 95.3 & 86.0 & 94.6 & 85.4 & 93.8 & 82.1\\
 & 16 &  & 94.5 & 88.1 & 94.7 & 88.3 & 94.4 & 87.0\\
 \hline
 & 8 &  & 94.5 & 78.1 & 94.1 & 78.2 & 94.3 & 75.9\\
$\tau^2=0.75$ & 12 &  & 95.4 & 87.7 & 95.5 & 87.2 & 94.0 & 83.3\\
 & 16 &  & 94.6 & 89.0 & 94.5 & 88.3 & 94.2 & 88.2\\
 \hline
 & 8 &  & 95.3 & 80.6 & 95.0 & 80.0 & 94.3 & 77.8\\
$\tau^2=1.0$  & 12 &  & 95.6 & 87.9 & 94.7 & 87.9 & 94.3 & 84.1\\
 & 16 &  & 94.5 & 89.0 & 94.2 & 88.5 & 95.1 & 88.7\\
\hline
\end{tabular}
\end{table}

\subsection{Example: screening test accuracy for alcohol problems}\label{sec:BMA-app}
\begin{figure}
\centering
\includegraphics[width=10cm]{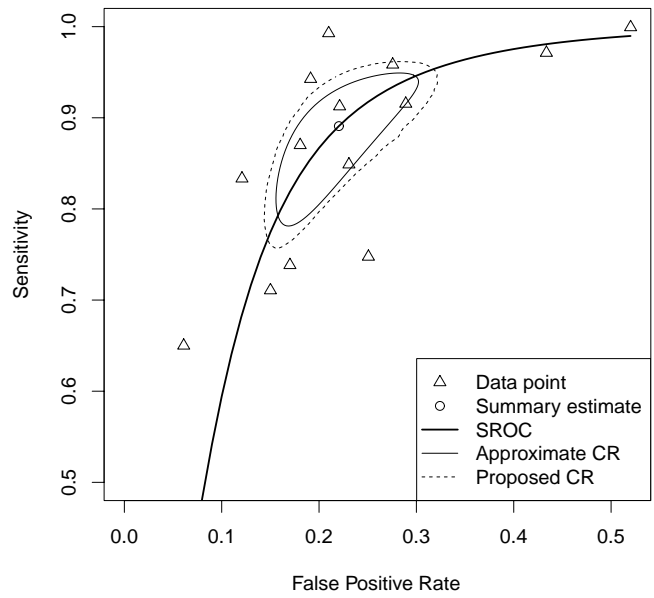}
\caption{Approximated and the proposed confidence regions (CRs) and summary receiver operating characteristics (SROC) curve.}
\label{fig:Audit}
\end{figure}

Here we provide a re-analysis of the dataset given in \cite{Kriston2008}, including $k=14$ studies regarding a short screening test for alcohol problems.
Following \cite{Reitsma2005}, we used logit-transformed values of sensitivity and specificity, denoted by $y_{Ai}$ and $y_{Bi}$, respectively, and associated standard errors $s_{Ai}$ and $s_{Bi}$.
For the bivariate summary data, we fitted the bivariate models (\ref{BMA1}) and (\ref{BMA2}), and computed $95\%$ CRs of $\mu$ based on the approximated CR of the form (\ref{NCR}) given in \cite{Reitsma2005}.
Moreover, we computed the proposed CR with $1000$ Monte Carlo samples for calculating $p$-values of the LRT, and $M=200$ evaluation points that were smoothed by a 7-point moving average for the CR boundary.
Following \cite{Reitsma2005}, the obtained two CRs of $(\mu_A,\mu_B)$ were transformed to the scale $({\rm logit}(\mu_A),1-{\rm logit}(\mu_B))$, where ${\rm logit}(\mu_A)$ and $1-{\rm logit}(\mu_B)$ are the sensitivity and false positive rate, respectively. 
The obtained two CRs are presented in Figure \ref{fig:Audit} with a plot of the observed data, summary points $\muh$, and the summary receiver operating curve. 
The approximate CR is smaller than the proposed CR, which may indicate that the approximation method underestimates the variability of estimating nuisance variance parameters.

\section{Network Meta-analysis}\label{sec:NMA}

\subsection{Multivariate random-effects model}
Suppose there are $p$ treatments in contract to a reference treatment, and let $y_{ir}$ be an estimator of relative treatment effect for the $r$th treatment in the $i$th study.
We consider the following multivariate random-effects model: 
\begin{equation}\label{NMA-ob}
y_i\sim  N(\th_i,S_i),  \ \ \ \ \  \th_i\sim  N(\beta,\Si), 
\ \ \ \ \ i=1,\ldots,k.
\end{equation}
where $y_i=(y_{i1},\ldots,y_{ip})^t$, $\th_i=(\th_{i1},\ldots,\th_{ip})^t$ is a vector of true treatment effects in the $i$th study, $\beta=(\beta_1,\ldots,\beta_p)^t$ is a vector of average treatment effects, and $S_i$ is the within-study variance-covariance matrix.
Here we focus on the model (\ref{NMA-ob}) known as the contrast-based model \citep{Salanti2008, Dias2016}, which is commonly used in practice.

In network meta-analysis, each study contains only $p_i(<p)$ treatments ($p_i$ usually ranges from 2 to 5); thereby, several components in $y_i$ cannot be defined.
When the corresponding treatments are not involved in the $i$th study, the corresponding components in $y_i$ and $S_i$ are shrunk to the sub-vector and sub-matrix, respectively, in the model (\ref{NMA-ob}).
Moreover, when the references treatment is not involved in the $i$th study, we can adopt the data argumentation approach of \cite{White2012}, in which a quasi-small data set is added to the reference arm, e.g. 0.001 events for 0.01 patients.  
To clarify the setting in which $y_i$ and $S_i$ are shrunk to the sub-vector and sub-matrix, respectively, we introduce an index $a_{ij}\in \{1,\ldots,p\}$, $j=1,\ldots,p_i$, representing the treatment estimates that are available in the $i$th study, and define the $p$-dimensional vector $x_{ij}$ of $0$'s, excluding the $a_{ij}$th element that is equal to $1$.
Moreover, we define $X_i=(x_{i1},\ldots, x_{ip_i})^t$, and $y_i$ and $S_i$ are the shrunken $p_i$-dimensional vector and $p_i\times p_i$ matrix of $y_i$ and $S_i$, respectively.
The model (\ref{NMA-ob}) can be rewritten as 
\begin{equation}\label{NMA1}
y_i\sim  N(X_i\th_i,S_i), \ \ \ \ \ \th_i\sim  N(\beta,\Si), \ \ \ \ i=1,\ldots,k.
\end{equation}
Regarding the structure of between study variance $\Sigma$, since there are rarely enough studies to identify the unstructured model of $\Sigma$, the compound symmetry structure $\Si=\tau^2P(0.5)$ is used in most cases \citep{White2015}, where $P(\rho)$ is a matrix with all diagonal elements equal to 1 and all off-diagonal elements equal to $\rho$.

We define $y=(y_1^t,\ldots,y_k^t)^t$, $X=(X_1^t,\ldots,X_k^t)^t$, $Z=\diag(X_1,\ldots,X_k)$, and $u=(u_1^t,\ldots,u_k^t)^t$ with $u_i\sim N(0,\tau^2P(0.5))$ independently for each $i$.
The hierarchical model (\ref{NMA1}) can be expressed as the following random-effects model:
\begin{equation}\label{NMA}
y=X\beta+Zu+\ep,
\end{equation}
where $\ep\sim N(0,S)$ with $S=\diag(S_1,\ldots,S_k)$.
The unknown parameters in (\ref{NMA}) are $\beta$ and $\tau^2$.
The log-likelihood of the model (\ref{NMA}) is given by
$$
L(\beta,\tau^2)=-\frac12\log|V(\tau^2)|-\frac12(y-X\beta)^tV(\tau^2)^{-1}(y-X\beta),
$$ 
where $V(\tau^2)=\tau^2 Z\{I_k\otimes P(0.5)\}Z^t+S$ and $\otimes$ denotes the Kronecker product.
Note that $y$ is an $N$-dimensional vector and $N=\sum_{i=1}^kp_i$ is the total number of comparisons.

\subsection{Confidence interval of the average treatment effect }
In network meta-analysis, we are interested in not only the average treatment effects $\beta_1,\ldots,\beta_p$ in contrast to the reference treatment, but also the treatment differences $\beta_j-\beta_k, j\neq k$.
To handle these issues in a unified manner, we focus on the linear combination $\eta=c^t\beta$ with known vector $c$.
Define a full-rank $p\times p$ matrix $A$ such that the first element of $A\beta$ is $\eta$.
The parameter $\eta$ is equivalent to $\beta_1$ when we use $XA^{-1}$ instead of $X$ in the model (\ref{NMA}), so that it is sufficient to consider a confidence interval of $\beta_1$.

Define $W_1$ and $W_2$ to be $N\times 1$ and $N\times (p-1)$ matrices such that $X=(W_1,W_2)$, and $\omega=(\beta_2,\ldots,\beta_p)^t$.
The model (\ref{NMA}) can be rewritten as
\begin{equation}\label{NMA-r}
y=W_1\beta_1+W_2\omega+Zu+\ep.
\end{equation}
We first consider testing of H$_0:\beta_1=\beta_{10}$, noting that $\omega$ and $\tau^2$ are nuisance parameters.
The LRT statistics can be defined as 
$$
T_{\beta_{10}}(Y)=\min_{\beta_1,\om,\tau^2}L(\beta_1,\om,\tau^2)-\min_{\omega,\tau^2}L(\beta_{10},\om,\tau^2),
$$  
where 
$$
L(\beta_1,\omega,\tau^2)
=\log |V(\tau^2)|+(y-W_1\beta_1-W_2\om)^tV(\tau^2)^{-1}(y-W_1\beta_1-W_2\om).
$$
Under H$_0$, the constrained maximum likelihood estimator $\omh_c$ and $\tah^2_c$ satisfy the following equations:
\begin{equation}\label{NMA-eq1}
\begin{split}
&W_2^tV(\tah^2_c)^{-1}r(\beta_{10},\omh_c)=0\\
&\tr\left\{V(\tah^2_c)^{-1}Q\right\}-r(\beta_{10},\omh_c)^tV(\tah^2_c)^{-1}QV(\tah^2_c)^{-1}r(\beta_{10},\omh_c)=0,
\end{split}
\end{equation}
where $r(\beta_1,\om)=y-W_1\beta_1-W_2\om$ and $Q=Z\{I_k\otimes P(0.5)\}Z^t$.
Under H$_0$, it holds that $y=W_1\beta_{10}+W_2\om+A(\tau^2)u$ for $u\sim N(0,I_N)$ and $A(\tau^2)$ is the Cholesky decomposition of $V(\tau^2)$ such that $A(\tau^2)A(\tau^2)^t=V(\tau^2)$.
The equation (\ref{NMA-eq1}) can be rewritten as 
\begin{equation}\label{NMA-eq2}
\begin{split}
&G_1(\omh_c,\tah^2_c,\om,\tau^2,u)
\equiv W_2^tV(\tah^2_c)^{-1}r(\om,\omh_c,\tau^2,u)=0\\
&G_2(\omh_c,\tah^2_c,\om,\tau^2,u)\\
& \ \ \ \ \ \ \equiv \tr\left\{V(\tah^2_c)^{-1}Q\right\}-r(\om,\omh_c,\tau^2,u)^tV(\tah^2_c)^{-1}QV(\tah^2_c)^{-1}r(\om,\omh_c,\tau^2,u)=0,
\end{split}
\end{equation}
with $r(\om,\omh_c,\tau^2,u)=W_2(\om-\omh_c)+A(\tau^2)u$.
The solution of the first equation $G_1(\omh_c,\tah^2_c,\om,\tau^2,u)=0$ with respect to $\om$ is given by 
\begin{equation*}
\om_{\ast}(u,\tau^2)=\omh_c-\{W_2^tV(\tah^2_c)^{-1}W_2\}^{-1}W_2^tV(\tah^2_c)^{-1}A(\tau^2)u.
\end{equation*}
By replacing $w$ with $\om_{\ast}(u,\tau^2)$ in the second equation in (\ref{NMA-eq2}), we obtain the following equation for $\tau^2$:
\begin{equation*}
\tr\left\{V(\tah^2_c)^{-1}Q\right\}-u^tA(\tau^2)^tB(\tah^2_c)V(\tah^2_c)^{-1}QV(\tah^2_c)^{-1}B(\tah^2_c)A(\tau^2)u=0,
\end{equation*}
where $B(\tau^2)=I_N-W_2\{W_2^tV(\tau^2)^{-1}W_2\}^{-1}W_2^tV(\tau^2)^{-1}$, and we define $\tau^2_{\ast}(u)$ be the solution of the above equation with respect to $\tau^2$.
Hence, the solutions of (\ref{NMA-eq2}) with respect to $\om$ and $\tau^2$ are given by $\om_{\ast}(u)=\om_{\ast}(u,\tau^2_{\ast}(u))$ and $\tau^2_{\ast}(u)$, respectively.
Concerning the weight function $w(U)$, from the implicit function theorem, it follows that 
$$
w(U)=\bigg|
\frac{\det\left(\partial G/\partial \omh_c^t,\partial G/\partial \tah^2\right)}
{\det\left(\partial G/\partial \om^t,\partial G/\partial \tau^2\right)}
\bigg|_{\om=\om_{\ast}(u), \tau^2=\tau^2_{\ast}(u)},
$$
where $G=(G_1^t,G_2)$.
From (\ref{NMA-eq2}), we have
\begin{align*}
&\frac{\partial G_1}{\partial \om^t}
=-\frac{\partial G_1}{\partial \omh_c^t}
=W_2^tV(\tah^2_c)^{-1}W_2\\
&\frac{\partial G_2}{\partial \om^t}
=-\frac{\partial G_2}{\partial \omh_c^t}
=-2W_2^tV(\tah^2_c)^{-1}QV(\tah^2_c)^{-1}\{W_2(\om-\omh_c)+A(\tau^2)u\}.
\end{align*}
On the other hand, because derivation of analytical expressions of the partial derivatives with respect to $\tau^2$ or $\tah_c^2$ requires tedious algebraic calculation, we can use numerical derivatives instead.
Therefore, we can carry out Algorithm 1 in Section \ref{sec:general} to compute the $p$-value of LRT, and the confidence interval can be obtained as well by inverting the LRT.

\subsection{Simulation study}
We investigate the performance of the proposed Monte Carlo (MC) method under practical network meta-analysis scenarios.
We compared the coverage probabilities of the MC method with those of widely used standard methods: the Wald-type confidence intervals based on REML estimates, the LR-based confidence interval.
Throughout the experiments, we set the nominal level $\alpha$ to $0.05$.
Following \cite{Noma2018}, we considered a quadrangular network comparing $4$ treatments (A, B, C, and D, regarding A as a reference).
The numbers of trials $k$ were set to $8, 12$ and $16$ and the detailed designs of trials are presented in Supplementary Material.   
To approximate practical situations of medical meta-analyses, we mimicked the simulation settings considered by \cite{Sidik2007}.
We first generated binomial data from $X_{ir}\sim {\rm Binomial}(n_{ir},p_{ir}), (i=1,\ldots,k)$, where $r = 0, 1, 2$, and $3$ corresponds to the treatments A, B, C, and D, respectively.
The response rate of treatment $A$, $p_{i0}$, was generated from a continuous uniform distribution on $[0.095,0.65]$ and we set $p_{ir}=p_{i0}\exp(\th_{ir})/\{1-p_{i0}+p_{i0}\exp(\theta_{ir})\}$ for $r=1,2$ and 3, which means that $\theta_{ir}$ is odds ratio (ORs) to the reference treatment A, i.e. $\theta_{ir}={\rm legit}(p_{ir})-{\rm legit}(p_{i0})$.
Also, the OR parameters $(\theta_{i1},\theta_{i2},\theta_{i3})$ were generated from a multivariate normal distribution $N(\mu,\tau^2P(0.5))$, where $\mu=(\mu_1,\mu_2,\mu_3)$ is a vector of the true average treatment effects set to $\mu=(0.4,0.7,1.0)$.
The sample sizes were set to equal one another, $n_{i0}=n_{i1}=n_{i2}=n_{i3}$ for any $i$ and were drawn from a discrete uniform distribution on 20 and 200.
From the generated binomial data $X_{ir}$'s, we calculated trial-specific summary statistics $y_i$ and $S_i$ in the standard manner \citep{Higgins2011}.
In the 2000 simulations, we evaluated empirical coverage probabilities for 95\% confidence intervals of the true parameters.
Due to the same computational reason as noted in Section \ref{sec:BMA-sim}, we only evaluated coverage rates of the confidence intervals.

The results of the simulations are shown in Table \ref{tab:NMA-sim}.
In general, the coverage probabilities of the REML confidence intervals are sightly better than the LR confidence intervals.
However, they showed undercoverage properties under moderate number of studies ($k=8, 12$) and large heterogeneity ($\tau=0.4$). 
On the there hand, the coverage probabilities of the proposed MC method were generally around the nominal level ($95\%$) in most cases.
Under the small number of studies $k=8$ and large heterogeneity ($\tau=0.4$), the coverage rates were relatively low, but even under these scenarios, they performed better than the ML and REML methods.

\begin{table}[!htb]
\caption{Simulated coverage probabilities (\%) for $95\%$ confidence intervals from the proposed Monte Carlo (MC) method, REML and LR. 
\label{tab:NMA-sim}
}
\centering
\vspace{0.5cm}
\begin{tabular}{ccccccccccccccccccc}
\hline
&&& \multicolumn{3}{c}{$\tau=0.2$} & \multicolumn{3}{c}{$\tau=0.3$} &\multicolumn{3}{c}{$\tau=0.4$}\\
$k$ & Methods && $\mu_1$ & $\mu_2$ &$\mu_3$ &$\mu_1$ &$\mu_2$ &$\mu_3$ &$\mu_1$ &$\mu_2$ &$\mu_3$ \\
 \hline
& LR &  & 93.0 & 92.6 & 91.8 & 91.4 & 90.6 & 91.1 & 88.2 & 88.4 & 88.6\\
8 & REML &  & 93.7 & 93.9 & 92.5 & 92.6 & 91.9 & 92.7 & 90.4 & 90.4 & 90.4\\
 & MC &  & 94.2 & 95.8 & 94.5 & 93.4 & 93.8 & 93.4 & 91.3 & 93.2 & 92.1\\
\hline
 & LR &  & 93.5 & 93.9 & 92.5 & 90.2 & 91.0 & 91.1 & 89.8 & 89.4 & 90.8\\
12 & REML &  & 94.1 & 94.6 & 93.4 & 91.7 & 92.0 & 92.3 & 91.3 & 90.6 & 92.2\\
 & MC &  & 95.3 & 96.1 & 94.6 & 93.4 & 93.8 & 93.9 & 92.7 & 92.7 & 92.6\\
 \hline
 & LR &  & 93.2 & 94.4 & 93.1 & 92.1 & 91.6 & 93.0 & 90.9 & 92.3 & 92.3\\
16 & REML &  & 93.9 & 95.0 & 93.7 & 92.9 & 92.4 & 93.5 & 92.3 & 92.8 & 93.2\\
 & MC &  & 95.0 & 95.7 & 94.5 & 93.7 & 94.1 & 94.2 & 92.8 & 94.1 & 93.9\\
\hline
\end{tabular}
\end{table}

\subsection{Example: Schizophrenia data}\label{sec:NMA-app}
\cite{Ades2010} carried out a network meta-analysis of antipsychotic medication for prevention of relapse of schizophrenia; this analysis includes $k=15$ trials comparing eight treatments with placebo.
In each trial, the outcomes available were the four outcome states at the end of follow-up: relapse, discontinuation of treatment due to intolerable side effects and other reasons, not reaching any of the three endpoints, and still in remission. 
We here considered the last outcome and adopted the odds ratio as the treatment effect measure.

We set the reference treatment to ``Placebo" and applied the multivariate random-effects model (\ref{NMA}).
The estimates of between-studies standard deviation $\tau$ were $0.28$ for the ML and $0.52$ for the REML estimation methods, respectively, which shows that there is substantial heterogeneity between studies.
In Table \ref{tab:SCZ} we present the results of three confidence intervals based on the MC method, the LR-based method with $p$-value calculated by the asymptotic distribution, and the REML method.
The number of Monte Carlo samples the MC method was consistently set to $10000$. 
In this analysis, the confidence intervals of MC were wider than those of LR.
On the other hand, REML produced wider intervals than MC in some treatments whereas REML produced narrower intervals than MC in the other treatments, which may be due to the difference between the ML and REML estimates of between study standard deviation.

\begin{table}[!htb]
\caption{Maximum likelihood (ML) and restricted maximum likelihood (REML) estimates of average treatment effects and confidence intervals from the proposed Monte Carlo (MC), likelihood ratio (LR) and REML methods in the application to network-meta analysis of schizophrenia data. 
\label{tab:SCZ}
}
\centering
\vspace{0.5cm}
\begin{tabular}{ccccccccccccccccccc}
\hline
Placebo v.s. & ML & MC & LR & REML & REML \\
\hline
Olanzapine & 4.91 & (2.30, 9.27) & (2.67, 8.55) & 4.52 & (2.15, 9.50) \\
Amisulpride & 3.38 & (1.30, 8.99) & (1.56, 7.30) & 3.36 & (1.22, 9.31) \\
Zotepine & 2.66 & (0.74, 9.29) & (0.91, 7.74) & 2.66 & (0.68, 10.32) \\
Aripiprazole & 2.07 & (0.57, 7.70) & (0.93, 4.58) & 2.07 & (0.67, 6.38) \\
Ziprasidone & 5.03 & (1.97, 12.38) & (2.32, 10.73) & 4.90 & (1.81, 13.26) \\
Paliperidone & 2.08 & (0.61, 7.53) & (0.90, 4.81) & 2.08 & (0.65, 6.67) \\
Haloperidol & 2.65 & (1.12, 5.46) & (1.26, 5.12) & 2.36 & (0.94, 5.95) \\
Risperidone & 5.46 & (2.12, 13.08) & (2.40, 11.82) & 5.05 & (1.74, 14.64) \\
\hline
\end{tabular}
\end{table}

\section{Discussions}\label{sec:Disc}
We developed a unified method for constructing confidence intervals of the average treatment effects in random-effects meta-analysis. 
The proposed confidence intervals are based on the LRT, and we proposed a Monte Carlo method to compute its $p$-value. 
In terms of specific applications, we discussed three types of meta-analysis, univariate meta-analysis, diagnostic meta-analysis, and network meta-analysis, and demonstrated the usefulness of the proposed method.
The R code for implementing the proposed methods together with applications to three datasets demonstrated in Sections \ref{sec:mag}, \ref{sec:BMA-app} and \ref{sec:NMA-app} are provided in Supplementary Material.
The developed inference method would be adapted to a variety of applications, e.g., the multivariate individual participant data meta-analysis \citep{Burke2016}. 
A limitation of the proposed methods might be rigorous justification of the exactness of the proposed inference methods.
Although the maximum likelihood estimator of the variance parameters are sufficient statistics in the case that all the within-study variances are the same, this property might not hold rigorously, under general conditions. 
Although there were no theoretical proofs concerning the sufficiency of this estimator under general conditions, but we could clearly demonstrate the proposed methods could provide almost exact confidence intervals in the simulation studies.

An alternative way to improve the coverage rates of confidence intervals is using Bayesian methods \citep[e.g.][]{Sutton2001}.
However, results from Bayesian methods may be sensitive to choices of prior distributions under the realistic number of studies as discussed in \cite{Lam2005}.
Also even if we use non-informative priors, frequentist validity of such Bayesian methods is generally  guaranteed under the large number of samples.
Therefore, we need to be careful for using Bayesian methods sine they do not necessarily work well in terms of accuracy of evidence synthesis.

In addition, the numerical results from our simulations and the illustrative examples suggest that statistical methods in the random-effects models should be selected carefully in practice. 
Historically, there have been many discrepant results between meta-analyses and subsequent large randomized clinical trials \citep{LeL1997}, and in these cases the meta-analyses have typically tended to provide false results as in the magnesium example in Section \ref{sec:mag}. 
Many systematic biases, for example, publication bias \citep{East1991} might be important sources of these discrepancies, but we should also be aware of the risk of providing overconfident and misleading interpretations caused by the statistical methods based on large sample approximations. 
Considering these risks, accurate inference methods would be preferred in practice. 
Although there have not been any accurate inference methods that can be broadly applied in random-effects meta-analyses, our approach in this article may provide an explicit solution to this relevant problem.

Finally, methodological research on extensions of random-effects meta-analyses to more complicated statistical models are still in progress \citep[e.g., multivariate network meta-analyses][]{Riley2017}, and the small sample problems generally exist in most of these applications. 
Our methods are applicable to these complicated models as well as more advanced approaches that might arise in future research. 
The developed methods should be effective tools as a unified methodological framework to obtain accurate solutions in medical evidence synthesis.

\section{Software}
R code used in this paper is available on github (https://github.com/sshonosuke/mcci-meta).

\section*{Acknowledgments}
This research was partially supported by JST CREST (grant number: JPMJCR1412) and JSPS KAKENHI (grant number: 17K19808, 15K15954, 18K12757).

\setcounter{table}{0}
\renewcommand{\thetable}{S\arabic{table}}
\newpage

\begin{center}
{\LARGE \bf Supplementary Material for ``A unified method for improved inference in random-effects meta-analysis" by Shonosuke Sugasawa and Hisashi Noma}
\end{center}

\section*{Tables}
We provide two tables showing the detailed results in the example of treatment of suspected acute myocardial infarction in Section 3.4 and the detailed designs of trials used in the simulation study of network meta-analysis in Section 5.3.

\begin{table}[htb!]
\caption{Estimates and confidence intervals of the average treatment effect of intravenous magnesium on myocardial infarction based on six methods in Section 3.4.
\label{tab:Mag}
}
\begin{center}
\begin{tabular}{ccccccccccccccccccc}
\hline
Method & Estimate & $95\%$ CI\\
\hline
MC & 0.449 & (0.149, 1.103)\\
KNHA & 0.438 & (0.200, 0.956)\\
DL & 0.448 & (0.233, 0.861)\\
LR & 0.449 & (0.191, 0.903)\\
REML & 0.438 & (0.213, 0.895)\\
PFE & 0.471 & (0.280, 0.791)\\
\hline
\end{tabular}
\end{center}

\caption{Numbers of trials for each study design included in the simulation studies in Section 5.3. 
\label{tab:NMA-sim-setting}
}
\centering
\vspace{0.2cm}
\begin{tabular}{ccccccccccccccccccc}
\hline
& $k=8$ & $k=12$ & $k=16$ \\
\hline
A vs. B & 1 & 2 & 2\\
A vs. C & 3 & 4 & 6\\
A vs. D & 1 & 2 & 3\\
B vs. C & -- & -- & 1\\
B vs. D & -- & 1 & 1\\
C vs. D & 1 & 1 & 1\\
A vs. C vs. D & 1 & 1 & 1 \\
B vs. C vs. D & 1 & 1 & 1\\
\hline
\end{tabular}
\end{table}

\end{document}